\documentclass{PoS}

\usepackage{amsmath, amssymb, amsfonts}
\usepackage{graphicx, dcolumn, bm}
\usepackage{psfrag}
\usepackage{epstopdf}
\usepackage{color}

\rightline{\sffamily RBRC-1052}\vspace*{-10mm}

\title{Neutral B meson mixing with static heavy and domain-wall light quarks}

\ShortTitle{
Neutral B meson mixing with static heavy and domain-wall light quarks}

\author{\speaker{Tomomi Ishikawa}\\
        RIKEN BNL Research Center, Upton, NY 11973, USA\\
        E-mail: \email{tomomi@quark.phy.bnl.gov}}

\author{Yasumichi Aoki\\
        Kobayashi-Maskawa Institute for the Origin of Particle and
	the Universe (KMI), Nagoya University, Nagoya 464-8602, Japan\\
	RIKEN BNL Research Center, Upton, NY 11973, USA\\
        E-mail: \email{yaoki@kmi.nagoya-u.ac.jp}}

\author{Taku Izubuchi\\
        Brookhaven National Laboratory, Upton, NY 11973, USA\\
	RIKEN BNL Research Center, Upton, NY 11973, USA\\
	E-mail: \email{izubuchi@quark.phy.bnl.gov}}

\author{Christoph Lehner\\
        Brookhaven National Laboratory, Upton, NY 11973, USA\\
	E-mail: \email{clehner@quark.phy.bnl.gov}}

\author{Amarjit Soni\\
        Brookhaven National Laboratory, Upton, NY 11973, USA\\
	E-mail: \email{soni@bnl.gov}}

\abstract{
Neutral B meson mixing matrix elements and B meson decay constants are
calculated.
Static approximation is used for b quark and domain-wall fermion formalism is
employed for light quarks.
The calculations are done on 2+1 flavor dynamical ensembles,
whose lattice spacings are $0.086$~fm and
$0.11$~fm with a fixed physical spatial volume of about $(2.7~{\rm fm})^3$.
In the static quark action, link-smearings are used to improve
the signal-to-noise ratio. 
We employ two kinds of link-smearings and their results  are combined
in taking a continuum limit.
For the matching between the lattice and the continuum theory,
one-loop perturbative calculations are used including
$O(a)$ improvements to reduce discretization errors.
We obtain SU(3) braking ratio $\xi=1.222(60)$ in the static limit of $b$ quark.
}

\FullConference{
31st International Symposium on Lattice Field Theory - LATTICE 2013\\
July 29 - August 3, 2013\\
Mainz, Germany}

\begin{document}

\section{Introduction}

One of the important purposes of flavor physics is an accurate
determination of the parameters of the Cabibbo-Kobayashi-Maskawa (CKM) matrix.
Now that existence of the Higgs particle was declared at the LHC experiments,
precise check of the CKM unitary triangle becomes even more important
for the search of New Physics
and B Physics provides valuable information to this effort.

The treatment of the $b$ quark is one of the challenging subjects in the
lattice QCD because of the multi-scale problem in which $b$ quark is
quite heavy ($\sim 4.2$~GeV), whereas $u$ and $d$ quarks are light (few MeV).
To resolve the difficulty, several approaches have been proposed and
useful results are beginning to be obtained.
Among them, Heavy Quark Effective Theory (HQET) is old-fashioned,
but a clean approach to this problem.
The static approximation is the lowest order of the HQET expansion
($1/m_Q$ expansion).
While the static approximation itself, without correction for $1/m_Q$ effects,
has $10\%$ level uncertainty,
its results are valuable as an anchor point
for the application to an interpolation from lower quark mass
region (charm quark mass region or higher).
Our current purpose is accurate calculations of the B Physics
quantities using static approximation for $b$ quark and chiral fermion
for light quarks.
While the $1/m_Q$ corrections definitely need to be addressed for
precision calculations,
the static approximation itself is more accurate for certain ratio quantities.
In the ratio the ambiguity from the static approximation is significantly
reduced by the suppression:
\begin{eqnarray}
O\left(\frac{\Lambda_{\rm QCD}}{m_{\rm b}}
\times\frac{m_{\rm s}-m_{\rm d}}{\Lambda_{\rm QCD}}
\right)\sim 2\%,
\end{eqnarray}
which makes the approximation competitive with other approaches
for one of the most important quantity in the $B^0-\overline{B}^0$ mixing
phenomena, SU(3) breaking ratio $\xi$.

In these proceedings we report our current status of the calculation
of $B$ meson decay constants and neutral $B$ meson mixing matrix
elements obtained using the static $b$ quark.

\section{Calculation}

\subsection{Action setup and ensembles}

As mentioned in the introduction,
the static approximation is used for $b$ quark.
In the static quark action, gluon link smearing is imposed to reduce
a notorious $1/a$ power divergence,
which is one of the key techniques in the HQET
suggested by Alpha Collaboration~\cite{DellaMorte:2003mn, Della Morte:2005yc}. 
We employ HYP1~\cite{Hasenfratz:2001hp} and HYP2~\cite{Della Morte:2005yc}
link smearing in this work.
For the light quark sector, we adopt the domain-wall fermion (DWF) formalism
to control the chiral symmetry, where the chiral symmetry plays an important
role to suppress unphysical operator mixing.

We use $2+1$ flavor dynamical DWF + Iwasaki gluon ensemble
generated by the RBC and UKQCD Collaborations~\cite{Aoki:2010dy},
listed in Tab.~\ref{TAB:ensembles}.
The pion masses $m_{\pi}$ at simulation points cover $290-420$~MeV range
and the finite size effects are modest so that
$m_{\pi}aL$ is not less than $4$.
\begin{table*}
\caption{\label{TAB:ensembles}
$2+1$ flavor dynamical DWF + Iwasaki gluon ensembles used in this calculation,
which is generated by the RBC and UKQCD Collaborations~\cite{Aoki:2010dy}.
In the table $m_{\rm l}$ and $m_{\rm h}$ represent $ud$ and $s$ quark mass
parameters, respectively, and $m_{\rm res}$ denotes a residual mass.
Physical $ud$ and $s$ quark mass parameters,
$m_{\rm ud, phys}$ and $m_{\rm s, phys}$, are obtained using
SU(2)$\chi$PT chiral fits.}
\begin{tabular}{ccccccc}
$\beta$ & $L^3\times T\times L_s$ & $a^{-1}$ [GeV] &
$am_{\rm ud, phys}$ & $am_{\rm s, phys}$ &
$am_{\rm res}$ & $m_{\rm l}/m_{\rm h}$ \\ \hline
$2.13$ & $24^3\times64\times16$ & $1.729(25)$ &
$0.00134(4)$ & $0.0379(11)$ &
$0.003152(43)$ & $0.005/0.04$ \\
        &                        &             & &
        &                &  $0.01/0.04$ \\ \hline
$2.25$ & $32^3\times64\times16$ & $2.280(28)$ &
$0.00100(3)$ & $0.0280(7)$ & $0.0006664(76)$ &
$0.004/0.03$ \\
        &                        &             & &
        &                 & $0.006/0.03$ \\
        &                        &             & &
        &                 & $0.008/0.03$ \\ \hline
\end{tabular}
\end{table*}

\subsection{Perturbative matching}\label{SEC:PT_matching}

We adopt perturbative matching,
where QCD and HEQT are matched in the continuum, then
the continuum and the lattice theory are matched in the HQET,
separately (two-step matching).
\begin{itemize}
\item {\bf Continuum matching}:
The QCD operators are renormalized in the $\overline{\rm MS}$(NDR) scheme
at a scale $\mu=m_{\rm b}$, $b$ quark mass scale.
The Fierz transformations in the arbitrary dimensions are specified
in the NDR scheme by Buras and Weisz~\cite{Buras:1989xd} where
evanescent operators are introduced.
The HQET operators are also renormalized in the $\overline{\rm MS}$(NDR)
scheme, and matched to the QCD at the $b$ quark mass scale.
The one-loop perturbative matching factors have been obtained
in Ref.~\cite{Eichten:1989zv} for quark bilinear operators
and in Refs.~\cite{Flynn:1990qz, Buchalla:1996ys} for $\Delta B=2$
four-quark operators.
\item {\bf Renormalization Group (RG) running}:
We use the RG running of the operators in the continuum HQET 
to go down to the lattice cut-off scale in order to avoid large logarithms
in the perturbation theory.
The two-loop anomalous dimensions have been calculated
in Refs.~\cite{Ji:1991pr, Broadhurst:1991fz} for quark bilinears and
in Refs.~\cite{Gimenez:1992is, Ciuchini:1996sr, Buchalla:1996ys}
for four-quark operators.
\item {\bf HQET matching}:
The matching in the HQET between the continuum and the lattice is made
at the lattice cut-off scale using one-loop perturbation theory
for our action setup,
in which $O(a)$ lattice discretization errors are taken into account and
the tad-pole improvement is used~\cite{Ishikawa:2011dd}.
\end{itemize}
We note that in the HQET matching, one might claim that $O(a)$ operators
can mix with $O(a^0)$ operators due to the $1/a$ power divergence,
causing the perturbative matching to fail.
The situation is, however, quite different from $O(1/m_Q)$ operators,
which are also higher dimensional ones and leave $O(\alpha_s^{l+1}/(m_Qa))$
uncertainty at $l^{\rm th}$-loop perturbation leading to huge error in taking
small $a$, because $\alpha_s$ only scales
logarithmically~\cite{Maiani:1991az}.
The $O(a)$ operators just bring $O(\alpha_s^{l+1})$ uncertainty at
$l^{\rm th}$-loop perturbation by mixing with $O(a^0)$ operators
keeping justification of the perturbation.

\subsection{Measurement}

\begin{figure*}
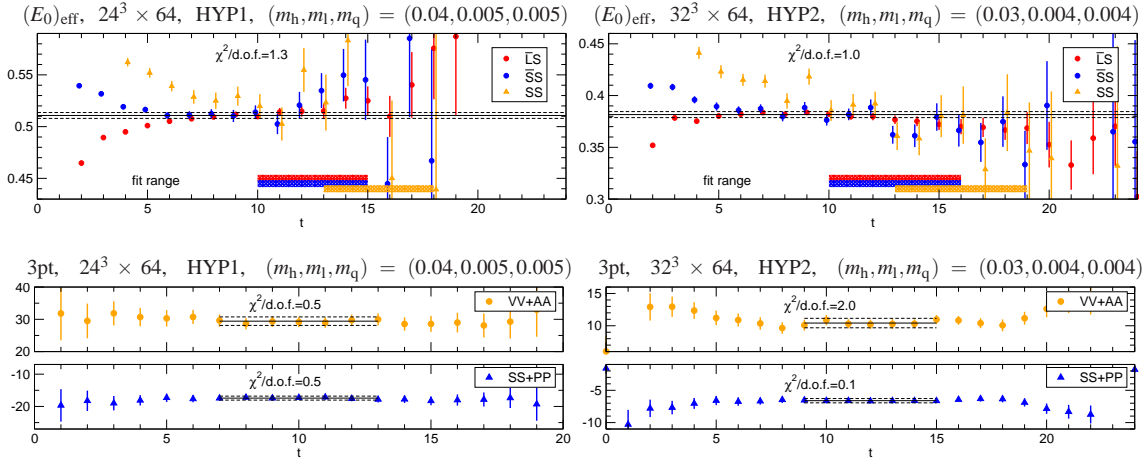

\begin{center}
\vspace*{-2mm}
\hspace*{-2mm}
\parbox{75mm}{
~{\scriptsize $(E_0)_{\rm eff}$, $24^3\times64$, HYP1,
$(m_{\rm h}, m_{\rm l}, m_{\rm q})=(0.04, 0.005, 0.005)$}
\includegraphics[scale=0.30, viewport = 0 0 705 250, clip]
{./Figures/Correlator_plots/24c64t_2PT_HYP1.eps}
\vspace*{-2mm}
}
\parbox{75mm}{
~{\scriptsize $(E_0)_{\rm eff}$, $32^3\times64$, HYP2,
$(m_{\rm h}, m_{\rm l}, m_{\rm q})=(0.03, 0.004, 0.004)$}
\includegraphics[scale=0.30, viewport = 0 0 705 250, clip]
{./Figures/Correlator_plots/32c64t_2PT_HYP2.eps}
\vspace*{-2mm}
}
\hspace*{-1mm}
\parbox{75mm}{
{\scriptsize ~3pt, $24^3\times64$, HYP1,
$(m_{\rm h}, m_{\rm l}, m_{\rm q})=(0.04, 0.005, 0.005)$}
\includegraphics[scale=0.30, viewport = 0 0 705 220, clip]
{./Figures/Correlator_plots/24c64t_3PT_HYP1.eps}
}
\parbox{75mm}{
{\scriptsize ~3pt, $32^3\times64$, HYP2,
$(m_{\rm h}, m_{\rm l}, m_{\rm q})=(0.03, 0.004, 0.004)$}
\includegraphics[scale=0.30, viewport = 0 0 705 220, clip]
{./Figures/Correlator_plots/32c64t_3PT_HYP2.eps}
}
\caption{
Examples of effective $E_0$ (two-point function)
and three-point function plots.}
\label{FIG:correlator_plots}
\end{center}
\end{figure*}

We use a gauge-invariant gaussian smearing for heavy and light quark fields
at source and sink, where the gaussian width is around $0.45$~fm.
We fit two-point and three-point functions:
\begin{eqnarray}
C_A^{\widetilde{L}S}(t, 0)&=&
\sum_{\vec{x}}\langle A_0(\vec{x}, t)A_0^{S\dag}(\vec{0}, 0)\rangle
\xrightarrow[t\gg0]{}
{\cal A}_A^{\widetilde{L}S}(e^{-E_0t}+e^{-E_0(T-t)}),\label{EQ:2pt-func_tLS}\\
C_A^{\widetilde{S}S}(t, 0)&=&
\sum_{\vec{x}}\langle A_0^S(\vec{x}, t)A_0^{S\dag}(\vec{0}, 0)\rangle
\xrightarrow[t\gg0]{}
{\cal A}_A^{\widetilde{S}S}(e^{-E_0t}+e^{-E_0(T-t)}),\label{EQ:2pt-func_tSS}\\
C_A^{SS}(t, 0)&=&
\langle A_0^S(\vec{0}, t)A_0^{S\dag}(\vec{0}, 0)\rangle
\xrightarrow[t\gg0]{}
{\cal A}_A^{SS}(e^{-E_0t}+e^{-E_0(T-t)}),\label{EQ:2pt-func_SS}
\end{eqnarray}
\begin{eqnarray}
C_L^{SS}(t_f, t, 0)&=&
\sum_{\vec{x}}
\langle A_0^S(\vec{0}, t_f)O_L(\vec{x}, t)A_0^{S\dag}(\vec{0}, 0)\rangle
\xrightarrow[t_f\gg t\gg0]{}
{\cal A}_L^{SS},\label{EQ:3pt-func_L}\\
C_S^{SS}(t_f, t, 0)&=&
\sum_{\vec{x}}
\langle A_0^S(\vec{0}, t_f)O_S(\vec{x}, t)A_0^{S\dag}(\vec{0}, 0)\rangle
\xrightarrow[t_f\gg t\gg0]{}
{\cal A}_S^{SS},\label{EQ:3pt-func_S}
\end{eqnarray}
where $A_0(\vec{x}, t)$ and $A_0^S(\vec{x}, t)$ are a local and 
a gaussian smeared axial heavy-light current, respectively.
$O_L$ and $O_S$ denote $\Delta B=2$ four-quark operators:
$O_L=[\overline{h}\gamma_{\mu}^Lq][\overline{h}\gamma_{\mu}^Lq]$,
$O_S=[\overline{h}P_Lq][\overline{h}P_Lq]$,
where $O_S$ comes into our calculation owing to a mixing with $O_L$ in
the HQET.
The correlator (\ref{EQ:2pt-func_SS}) is noisy, since volume summation
at sink is not taken.
Nevertheless, we need this correlator for extracting matrix elements
from three-point functions due to a reason specific to static quark,
in which energies of states in the static limit do not depend on their momentum
~\cite{Christ:2007cn}.
The correlator fitting is made simultaneously for three two-point functions
(\ref{EQ:2pt-func_tLS}), (\ref{EQ:2pt-func_tSS}) and (\ref{EQ:2pt-func_SS}), 
while separately for two three-point functions (\ref{EQ:3pt-func_L}) and
(\ref{EQ:3pt-func_S}).
The examples of the effective $E_0$ and three-point function plots are
presented in Fig.~\ref{FIG:correlator_plots}, in which fit results are shown.

After fitting the correlators, B meson decay constants $f_B$ and mixing
matrix elements ${\cal M}_B$ are obtained by:
\begin{eqnarray}
f_B=\frac{\Phi_B}{\sqrt{m_B}}
=\sqrt{\frac{2}{m_B{\cal A}_A^{\widetilde{S}S}}}
Z_A{\cal A}_A^{\widetilde{L}S},\;\;\;
{\cal M}_B=m_BM_B
=
\frac{2m_B}{{\cal A}_A^{SS}e^{-E_0t_f}}
(Z_L{\cal A}_L^{SS}+Z_S{\cal A}_S^{SS}),
\end{eqnarray}
where $Z_A$, $Z_L$ and $Z_S$ are matching factors between continuum QCD
and lattice HQET calculated in Ref.~\cite{Ishikawa:2011dd}.
Note that operators $A_0(\vec{x}, t)$, $O_L(\vec{x},t)$ and $O_S(\vec{x},t)$
are all $O(a)$ improved using one-loop perturbation~\cite{Ishikawa:2011dd}.

\subsection{Chiral and continuum extrapolations}

In the chiral and continuum extrapolation, we basically adopt
SU(2) heavy meson chiral perturbasion theory (SU(2)HM$\chi$PT).
(For the detailed expressions, see Ref.~\cite{Albertus:2010nm}.)
As discussed in Ref.~\cite{Allton:2008pn}, SU(2)$\chi$PT fit does not
converge in the pion mass region above $420$ MeV.
Our simulation points are below that mass.
On current statistics, a linear fit function hypothesis cannot be excluded,
then we take an average of results from SU(2)$\chi$PT and linear fit
as a central value.
(There is no distinction of the fit function between SU(2)$\chi$PT and linear
for $B_s$ quantities.)
Other thing we should mention is that we have only one sea $s$ quark
mass for each ensemble and is off from the physical point
as shown in Tab.~\ref{TAB:ensembles},
which is a source of error in the SU(2)$\chi$PT fits.
We estimate this error using SU(3)$\chi$PT as a model.

\begin{figure}
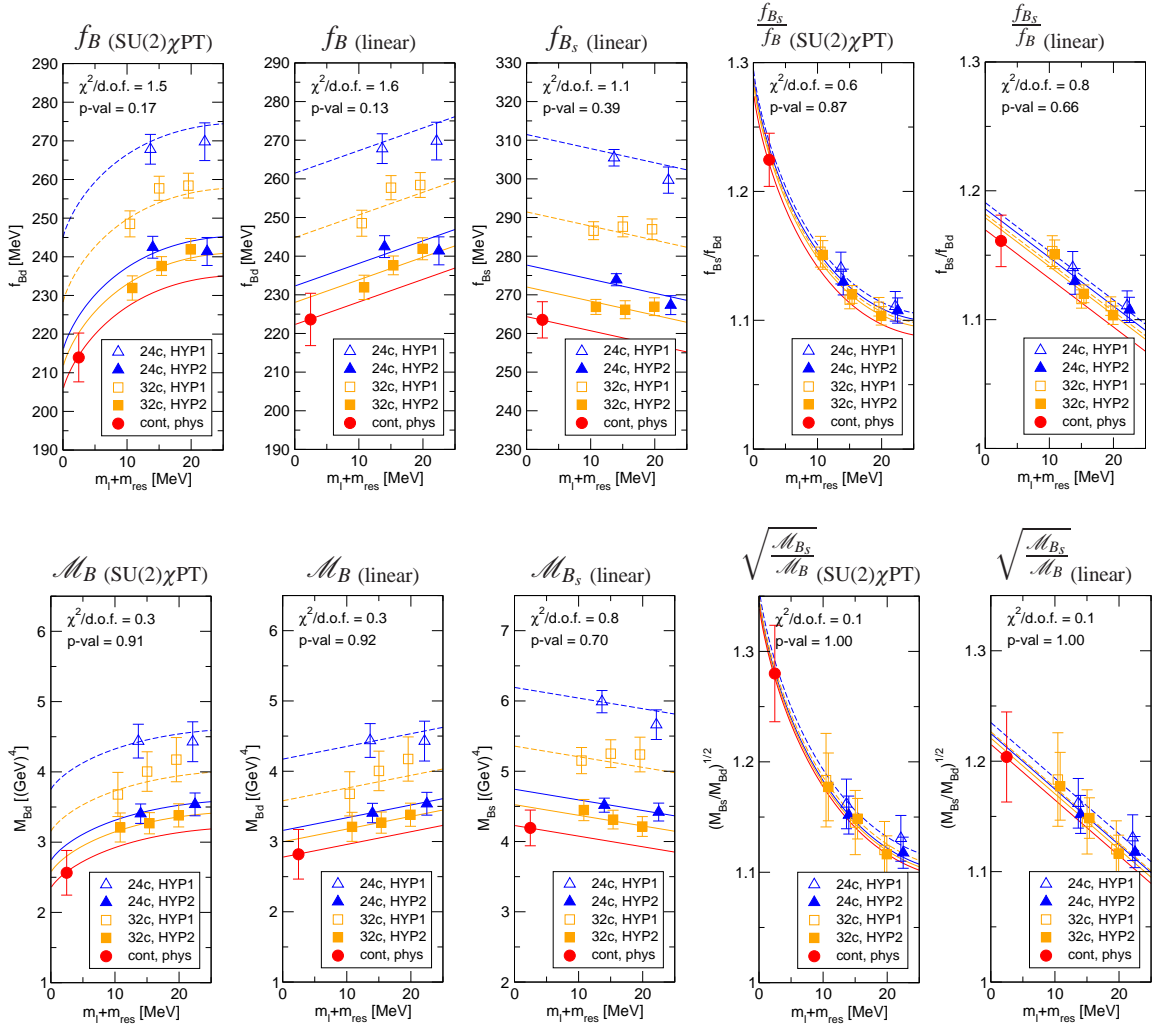

\begin{center}
\parbox{28mm}{
\hspace*{+9mm}{$f_{B~\mbox{\scriptsize (SU(2)$\chi$PT)}}$}
\includegraphics[scale=0.35, viewport = 0 0 240 470, clip]
{./Figures/Chiral_fit/fBd_SU2/plot.eps}
}
\hspace*{+1mm}
\parbox{28mm}{
\hspace*{+11mm}{$f_{B~\mbox{\scriptsize (linear)}}$}
\includegraphics[scale=0.35, viewport = 0 0 240 470, clip]
{./Figures/Chiral_fit/fBd_linear/plot.eps}
}
\hspace*{+1mm}
\parbox{28mm}{
\hspace*{+10mm}{$f_{B_s~\mbox{\scriptsize (linear)}}$}
\includegraphics[scale=0.35, viewport = 0 0 240 470, clip]
{./Figures/Chiral_fit/fBs_SU2/plot.eps}
}
\hspace*{+1mm}
\parbox{28mm}{
\vspace*{-3mm}
\hspace*{+7mm}{$\frac{f_{B_s}}{f_B}_{\mbox{\scriptsize (SU(2)$\chi$PT)}}$}
\includegraphics[scale=0.35, viewport = 0 0 240 470, clip]
{./Figures/Chiral_fit/fBr_SU2/plot.eps}
}
\hspace*{+1mm}
\parbox{28mm}{
\vspace*{-3mm}
\hspace*{+10mm}{$\frac{f_{B_s}}{f_B}_{\mbox{\scriptsize (linear)}}$}
\includegraphics[scale=0.35, viewport = 0 0 240 470, clip]
{./Figures/Chiral_fit/fBr_linear/plot.eps}
}
\end{center}
\begin{center}
\parbox{28mm}{
\vspace*{+4mm}
\hspace*{+6mm}{${\cal M}_{B~\mbox{\scriptsize (SU(2)$\chi$PT)}}$}
\includegraphics[scale=0.35, viewport = 0 0 240 470, clip]
{./Figures/Chiral_fit/MBd_SU2/plot.eps}
}
\hspace*{+1mm}
\parbox{28mm}{
\vspace*{+4mm}
\hspace*{+9mm}{${\cal M}_{B~\mbox{\scriptsize (linear)}}$}
\includegraphics[scale=0.35, viewport = 0 0 240 470, clip]
{./Figures/Chiral_fit/MBd_linear/plot.eps}
}
\hspace*{+1mm}
\parbox{28mm}{
\vspace*{+4mm}
\hspace*{+8mm}{${\cal M}_{B_s~\mbox{\scriptsize (linear)}}$}
\includegraphics[scale=0.35, viewport = 0 0 240 470, clip]
{./Figures/Chiral_fit/MBs_SU2/plot.eps}
}
\hspace*{+1mm}
\parbox{28mm}{
\hspace*{+5mm}{$\sqrt{\frac{{\cal M}_{B_s}}{{\cal M}_B}}_{\mbox{\scriptsize (SU(2)$\chi$PT)}}$}
\includegraphics[scale=0.35, viewport = 0 0 240 470, clip]
{./Figures/Chiral_fit/MBr_SU2/plot.eps}
}
\hspace*{+1mm}
\parbox{28mm}{
\hspace*{+8mm}{$\sqrt{\frac{{\cal M}_{B_s}}{{\cal M}_B}}_{\mbox{\scriptsize (linear)}}$}
\includegraphics[scale=0.35, viewport = 0 0 240 470, clip]
{./Figures/Chiral_fit/MBr_linear/plot.eps}
}
\end{center}
\caption{Chiral/continuum fit of B meson decay constants and $\Delta B=2$ mixing
 matrix elements.}
\label{FIG:Chiral_fit}
\end{figure}

\section{Results and future perspective}

We obtain chiral and continuum extrapolated (preliminary) results:
\begin{eqnarray}
f_B=219(19)(26)~[{\rm MeV}],\;\;\;
f_{B_s}=264(19)(32)~[{\rm MeV}],\;\;\;
f_{B_s}/f_B=1.193(42)(26),\\
{\cal M}_B=2.69(47)(32)~[({\rm GeV})^4],\;\;\;
{\cal M}_{B_s}=4.19(50)(50)~[({\rm GeV})^4],\;\;\;
\xi=1.222(60)(27),
\end{eqnarray}
where the first error denotes statistical and systematic errors including:
(1)~chiral/continuum extrapolation, (2)~finite volume effect,
(3)~one-loop $O(a)$ improvement error, (4)~one-loop renormalization error,
(5)~$g_{B^{\ast}B\pi}$ error, (6)~scale ambiguity,
(7)~physical quark mass error, (8)~off-physical $s$ quark mass ambiguity,
(9)~fit range ambiguity.
The second error shows static approximation ambiguity estimated by
$\Lambda_{\rm QCD}/m_{\rm b}$ with $\Lambda_{\rm QCD}\sim 0.5~{\rm GeV}$ and
$m_{\rm b}\sim 4.2~{\rm GeV}$.
\begin{figure}
\begin{center}
\includegraphics[scale=0.40, viewport = 0 0 880 270, clip]
{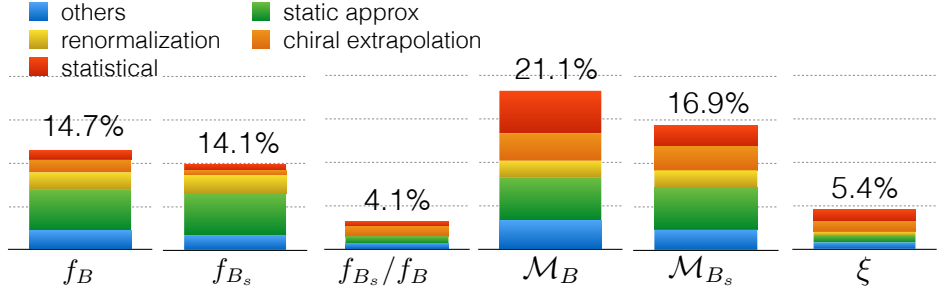}
\end{center}
\vspace*{-3mm}
\caption{Current error budget for physical quantities.
The numbers show total relative errors.}
\label{FIG:Error_budget}
\end{figure}

Fig.~\ref{FIG:Error_budget} shows error budgets for B meson decay constants,
mixing matrix elements and their ratio quantities.
Currently, statistical error and uncertainty from chiral extrapolation is
significantly large in the matrix element sector compared with decay constants.
The static approximation error in non-ratio quantities occupies large portion
of the total uncertainty, which is the potential limitation of
the approximation.
To reduce current large uncertainties, we are making following improvements:
\begin{itemize}
\item {\bf All-Mode-Averaging(AMA)}:
Using an error reduction technique proposed in Ref.~\cite{Blum:2012uh},
it is possible to make the statistical error significantly small.
The calculation using AMA is on-going on the same ensemble of this work.
In the case of the $32^3\times64$ ensemble and the lightest light quark
mass parameter, we use a deflated CG with $130$ low-mode eigenvectors,
relax the stopping condition of the CG from $10^{-8}$ to $3\times10^{-3}$ 
as an approximation and put $64$ sources, then the current status shows
$2-4$ times more efficiency compared with a deflated CG without AMA.
M\"{o}bius domain-wall fermion (MDWF) is also applicable for the approximation,
giving another factor of $2$ gain.
\item {\bf On physical light quark mass point simulation}:
The RBC and UKQCD Collaborations has been generating $48^3\times96\times24$ and
$64^3\times128\times16$ ensemble at almost physical pion mass
$m_{\pi}\sim135~{\rm MeV}$ using MDWF~\cite{Blum:2013lattice}.
Calculations on these ensemble enable us to completely remove
the chiral extrapolation uncertainty.
\item {\bf Non-perturbative renormalization}:
One-loop renormalization error is estimated to be $5\%$ for decay constants
and matrix elements, $0\%$ for $f_{B_s}/f_B$ and $1\%$ for $\xi$.
Apparently non-perturbative renormalization is necessary for non-ratio
quantities.
The renormalization would be made by RI-MOM scheme with
an additional renormalization condition due to the $1/a$ power divergence.
\item {\bf Beyond static approximation}:
It is apparent that the inclusion of $1/m_Q$ correction is necessary
to get out of the static approximation for high precision results.
Even when the results in the static limit are used for the interpolation
with lower quark mass region, the information of the $1/m_Q$ slope
would be important.
The matching should be performed non-perturbatively, otherwise it is
theoretically incorrect (no continuum limit), as depicted
in Sec.~\ref{SEC:PT_matching}.
To perform the non-perturbative matching, step scaling technique is
needed, where HQET is first matched to QCD on a super-fine lattice with small
volume, then the HQET at lower-energy with large volume is achieved
by the step scaling~\cite{Heitger:2003nj}.
\end{itemize}
While computationally challenging,
these improvements would give substantial impact
on the high precision B physics.

\section*{Acknowledgements}

Computations of observables in this work were performed on
QCDOC computers at RIKEN-BNL Research Center (RBRC) and
Brookhaven National Laboratory (BNL),
RIKEN Integrated Cluster of Clusters (RICC) at RIKEN, Wako,
KMI computer $\varphi$ at Nagoya University
and resources provided by the USQCD Collaboration
funded by the U.S. Department of Energy.
Y.~A. is supported by the JSPS Kakenhi Grant Nos.~21540289 and 22224003.
The work of C.~L, T.~Izubuchi and A.~S is supported in part
by the US DOE contract No.~DE-AC02-98CH10886.


\begin{thebibliography}{99}

\bibitem{DellaMorte:2003mn} 
  M.~Della Morte {\it et al.}, 
  Phys.\ Lett.\ B {\bf 581}, 93 (2004)
  [hep-lat/0307021].


\bibitem{Della Morte:2005yc}
  M.~Della Morte, A.~Shindler and R.~Sommer,
  JHEP {\bf 0508}, 051 (2005)
  [arXiv:hep-lat/0506008].


\bibitem{Hasenfratz:2001hp}
  A.~Hasenfratz and F.~Knechtli,
  Phys.\ Rev.\  D {\bf 64}, 034504 (2001)
  [arXiv:hep-lat/0103029].


\bibitem{Aoki:2010dy} 
  Y.~Aoki {\it et al.}, 
  Phys.\ Rev.\ D {\bf 83}, 074508 (2011)
  [arXiv:1011.0892 [hep-lat]].


\bibitem{Buras:1989xd}
  A.~J.~Buras and P.~H.~Weisz,
  Nucl.\ Phys.\  B {\bf 333} (1990) 66.


\bibitem{Eichten:1989zv}
  E.~Eichten and B.~R.~Hill,
  Phys.\ Lett.\  B {\bf 234}, 511 (1990).


\bibitem{Flynn:1990qz}
  J.~M.~Flynn, O.~F.~Hernandez and B.~R.~Hill,
  Phys.\ Rev.\  D {\bf 43}, 3709 (1991).


 \bibitem{Buchalla:1996ys}
  G.~Buchalla,
  Phys.\ Lett.\  B {\bf 395}, 364 (1997)
  [arXiv:hep-ph/9608232].


\bibitem{Ji:1991pr}
  X.~D.~Ji and M.~J.~Musolf,
  Phys.\ Lett.\  B {\bf 257}, 409 (1991).


\bibitem{Broadhurst:1991fz}
  D.~J.~Broadhurst and A.~G.~Grozin,
  Phys.\ Lett.\  B {\bf 267}, 105 (1991)
  [arXiv:hep-ph/9908362].


\bibitem{Gimenez:1992is}
  V.~Gimenez,
  Nucl.\ Phys.\  B {\bf 401}, 116 (1993).


\bibitem{Ciuchini:1996sr}
  M.~Ciuchini, E.~Franco and V.~Gimenez,
  Phys.\ Lett.\  B {\bf 388}, 167 (1996)
  [arXiv:hep-ph/9608204].


\bibitem{Ishikawa:2011dd} 
  T.~Ishikawa {\it et al.}, 
  JHEP {\bf 1105}, 040 (2011)
  [arXiv:1101.1072 [hep-lat]].


\bibitem{Maiani:1991az} 
  L.~Maiani, G.~Martinelli and C.~T.~Sachrajda,
  Nucl.\ Phys.\ B {\bf 368}, 281 (1992).


\bibitem{Christ:2007cn} 
  N.~H.~Christ, T.~T.~Dumitrescu, O.~Loktik and T.~Izubuchi,
  PoS LAT {\bf 2007}, 351 (2007)
  [arXiv:0710.5283 [hep-lat]].


\bibitem{Albertus:2010nm} 
  C.~Albertus {\it et al.}, 
  Phys.\ Rev.\ D {\bf 82}, 014505 (2010)
  [arXiv:1001.2023 [hep-lat]].


\bibitem{Allton:2008pn} 
  C.~Allton {\it et al.},  
  Phys.\ Rev.\ D {\bf 78}, 114509 (2008)
  [arXiv:0804.0473 [hep-lat]].


\bibitem{Blum:2012uh} 
  T.~Blum, T.~Izubuchi and E.~Shintani,
  Phys.\ Rev.\ D {\bf 88}, 094503 (2013)
  [arXiv:1208.4349 [hep-lat]].


\bibitem{Blum:2013lattice}
  T.~Blum {\it et al.},
  PoS LATTICE {\bf 2013}, 404 (2013) (in these proceedings).


\bibitem{Heitger:2003nj} 
  J.~Heitger {\it et al.}, 
  JHEP {\bf 0402}, 022 (2004)
  [hep-lat/0310035].


\end{thebibliography}
\end{document}